\documentstyle[12pt]{article}
\newlength{\pubnumber} 

\catcode`\@=11
\@addtoreset{equation}{section}
\def\section{\@startsection{section}{1}{\z@}{3.5ex plus 1ex minus .2ex}
 {2.3ex plus .2ex}{\large\bf}}
\def\subsection{\@startsection{subsection}{2}{\z@}{2.3ex plus .2ex}
 {2.3ex plus .2ex}{\bf}}

\begin{document}

\begin{titlepage}
\samepage{
\setcounter{page}{1}
\rightline{IASSNS-HEP-95/12}
\rightline{McGill/95-10}
\rightline{\tt hep-th/9503055}
\rightline{published in {\it Phys.\ Rev.\ Lett.}\/ {\bf 74} (1995) 4767}
\rightline{February 1995}
\vfill
\begin{center}
 {\Large \bf String Theory, Misaligned Supersymmetry,\\
  and the Supertrace Constraints\\}
\vfill
 {\large Keith R. Dienes$^1$\footnote{E-mail address:
  dienes@sns.ias.edu.}, Moshe Moshe$^2$\footnote{E-mail address:
  phr74mm@vmsa.technion.ac.il}, and Robert C. Myers$^3$\footnote{
  E-mail address:  rcm@hep.physics.mcgill.ca}\\}
\vspace{.25in}
 {\it  $^1$  School of Natural Sciences, Institute for Advanced Study\\
  Olden Lane, Princeton, NJ  ~08540~  USA\\}
\vspace{.05in}
 {\it $^2$ Department of Physics, Technion -- Israel Inst.\ of Technology\\
  Haifa 32000, Israel\\}
\vspace{.05in}
 {\it $^3$ Department of Physics, McGill University \\
  3600 University St., Montr\'eal, Qu\'ebec  H3A-2T8  Canada\\}
\end{center}
\vfill
\begin{abstract}
  {\rm We demonstrate that string consistency in four spacetime dimensions
   leads to a spectrum of string states which satisfies the
   supertrace constraints ${\rm Str}\,{\bf 1}=0$ and
   ${\rm Str}\,M^2 \propto \Lambda$ at tree level,
   where $\Lambda$ is the one-loop string cosmological constant.
   This result holds for a large class of string theories, including
   critical heterotic strings.
   For strings lacking spacetime supersymmetry,
   these supertrace constraints will be satisfied as a consequence of
   a hidden ``misaligned supersymmetry'' in the string spectrum.
   These results thus severely constrain the possible supersymmetry-breaking
   scenarios in string theory, and suggest a new intrinsically stringy
   mechanism whereby such supertrace constraints may be satisfied without
   phenomenologically unacceptable consequences.}
\end{abstract}

\vfill}
\end{titlepage}

\setcounter{footnote}{0}

\def\beq{\begin{equation}}
\def\eeq{\end{equation}}
\def\beqn{\begin{eqnarray}}
\def\eeqn{\end{eqnarray}}

\def\half{{\textstyle{1\over 2}}}
\def\third{{\textstyle {1\over3}}}
\def\quarter{{\textstyle {1\over4}}}
\def\field{{{\rm field}}}
\def\strin{{{\rm string}}}

\def\Str{{{\rm Str}\,}}
\def\qbar{{\overline{q}}}
\def\calF{{\cal F}}
\def\bone{{\bf 1}}
\def\bZ{{\bf Z}}

\def\cals{{\cal S}}
\def\calf{{\cal F}}
\def\calm{ {\cal M} }
\def\tautwo{{ \tau_2}}
\def\tauone{{ \tau_1}}
\def\qbar{{  \overline{q} }}
\def\Z{{ \bf Z}}

\def\inbar{\,\vrule height1.5ex width.4pt depth0pt}

\def\IC{\relax\hbox{$\inbar\kern-.3em{\rm C}$}}
\def\IQ{\relax\hbox{$\inbar\kern-.3em{\rm Q}$}}
\def\IR{\relax{\rm I\kern-.18em R}}
 \font\cmss=cmss10 \font\cmsss=cmss10 at 7pt
\def\IZ{\relax\ifmmode\mathchoice
 {\hbox{\cmss Z\kern-.4em Z}}{\hbox{\cmss Z\kern-.4em Z}}
 {\lower.9pt\hbox{\cmsss Z\kern-.4em Z}}
 {\lower1.2pt\hbox{\cmsss Z\kern-.4em Z}}\else{\cmss Z\kern-.4em Z}\fi}

\hyphenation{su-per-sym-met-ric non-su-per-sym-met-ric}
\hyphenation{space-time-super-sym-met-ric}
\hyphenation{mod-u-lar mod-u-lar--in-var-i-ant}

\setcounter{footnote}{0}

In quantum field theories with broken supersymmetry, the divergence properties
of
amplitudes are governed by the values of various supertraces calculated over
the particles
in the resulting spectrum.
For example, in four-dimensional spacetime,
$\Str M^4$ controls the logarithmic divergences in the vacuum energy density,
while $\Str M^2$ and $\Str M^0\equiv \Str {\bf 1}$
control the quadratic and quartic divergences respectively.
If the supersymmetry (SUSY) is unbroken, each of these supertraces of course
vanishes as
a consequence of strict level-by-level degeneracies between bosonic and
fermionic
degrees of freedom.
It is phenomenologically important, however, to construct non-SUSY field
theories
which retain the soft divergence behavior of their SUSY counterparts,
hopefully cancelling the quartic and quadratic divergences which might appear.
As is well-known, this can be achieved at tree level by breaking the SUSY
either
spontaneously, or through the addition of certain ``soft'' breaking terms;
indeed, in many
cases the vanishing of $\Str {\bf 1}$ and $\Str M^2$ is preserved.
The problem with these scenarios, however, is that they satisfy these
two constraints in a multiplet-by-multiplet fashion, so
that the mass of each state in the
broken theory is constrained to be relatively close to that of its former
superpartner.
Since this is unacceptable from a phenomenological standpoint, one must
therefore
rely on further quantum effects in order to lift these constraints.
One then finds that although $\Str \bone$ continues to vanish,
$\Str M^2$ takes a non-zero, model-dependent value.

In this paper we consider the corresponding situation in string theory, and
find that
at tree level, the general requirements of string consistency lead to similar
supertrace constraints.  Specifically, defining our string-theoretic
supertraces as
\beq
      \Str M^{2\beta}~\equiv~ \lim_{\gamma\to 0}\,
         \left\lbrace\sum_{\rm states}\, (-1)^F\, (M_i)^{2\beta}
        \,e^{-\gamma M_i^2}\right\rbrace~,
\label{regulator}
\eeq
we find that for a large class of tachyon-free string theories in four
dimensions,
\beq
         \Str \bone ~=~ 0 ~~~~~   {\rm and} ~~~~~
           \Str M^2 ~=~ -{3\over 4\pi^2 }\, \Lambda_\strin
\label{supertraces}
\eeq
where $\Lambda_\strin$ is the corresponding (finite) one-loop string-theoretic
cosmological constant.
Thus, the spacetime bosons and fermions at all string mass levels
must always arrange themselves at tree level so that these two supertrace
constraints
are satisfied.
Unlike the case in field theory, however,
we will find that these results rely
on only the general properties of string consistency (in particular,
the presence of modular invariance and the absence of physical tachyons).
These results are therefore independent of the particular string model
in question,
and consequently have broader applicability
than in field theory.

A second and perhaps more important difference
concerns the {\it manner}\/ in which these constraints are satisfied.
In string theories with spacetime SUSY, $\Lambda_\strin=0$ and
these constraints are trivially satisfied through an exact boson/fermion
degeneracy.
This is just as in the field theory case.
However, for string theories {\it without}\/ spacetime SUSY,
these constraints need not be satisfied multiplet-by-multiplet.  Rather, as we
will
discuss, these constraints are generally satisfied in a different manner,
through
a so-called ``misaligned SUSY''.
Misaligned SUSY therefore represents an entirely new stringy scenario
whereby constraints such as those in Eq.~(\ref{supertraces}) may be satisfied
without phenomenologically
unacceptable consequences.  As we shall see, this alternative scenario is
possible
in string theory because of the existence of an {\it infinite}\/ tower of
string states,
thereby permitting the freedom to satisfy the supertrace constraints across the
entire string
spectrum, rather than multiplet-by-multiplet.  Indeed,
in non-SUSY string models it is not necessary (or often even possible)
to make reference to a (broken) multiplet structure in the spectrum.
The bosonic and fermionic states which appear will nevertheless conspire
to exhibit a ``misaligned SUSY'' and satisfy Eq.~(\ref{supertraces}) in a
highly non-trivial manner.

Let us begin by first reviewing the appearance of the supertrace constraints in
field theory.
Perhaps the simplest manner in which they arise is through the calculation of
the field-theoretic vacuum energy density (cosmological constant), given to
lowest order as
\beqn
   \Lambda_{\rm field}
   &=& \phantom{-}\half\, \sum_i \,(-1)^F\, \int {d^D p \over (2\pi)^D}
\,\log(p^2+M_i^2)\nonumber\\
   &=& -\half\, \sum_i \,(-1)^F
  \int {d^D p \over (2\pi)^D} \int_0^\infty {dt\over t}
\,e^{-(p^2+M_i^2)t}\nonumber\\
   &=& -{1\over 2} {1\over (4\pi)^{D/2}} \sum_i\, (-1)^F
  \,\int_0^\infty {dt\over t^{1+D/2}} \, e^{-M_i^2 t}~.
\label{Lambdadef}
\eeqn
Here the summations are over all states in the theory (with corresponding
masses $M_i$),
and we have kept the spacetime dimension $D$ arbitrary.
In the second line we have passed to a Schwinger proper-time representation
wherein any ultraviolet divergences from $p^\mu\to\infty$ appear
as a divergence as $t\to 0$, while infrared divergences appear as $t\to\infty$.

We are concerned with the divergence properties of $\Lambda_{\rm field}$,
and from Eq.~(\ref{Lambdadef}) these can now easily be determined.
The absence of any infrared divergence from the $t\to\infty$ region
is guaranteed if there are no tachyonic states with $M_i^2<0$.
Ultraviolet divergences, on the other hand, would appear as $t\to 0$.
These will therefore be absent if
\beq
   \sum_i \,(-1)^F\,  e^{-M_i^2 t}~\sim~ t^\alpha ~~{\rm with}~~\alpha > D/2~
\label{conditionone}
\eeq
as $t\to 0$.
Since we are in the $t\to 0$ limit, we can expand the exponential,
$e^{-M_i^2 t} = 1 - M_i^2 t + M_i^4 t^2/2 + ...$,
and thereby obtain the separate supertrace conditions
$       \Str M^{2\beta} \equiv \sum_i (-1)^F (M_i)^{2\beta} = 0,
$
valid for $\beta=0,1,..., [D/2]$
where $[x]$ denotes the greatest integer less than or equal to $x$.
In particular, for $D=4$, this yields the three separate supertrace constraints
          $\Str \bone = \Str M^2 = \Str M^4 = 0$,
with logarithmic, quadratic, and quartic divergences respectively if the $\Str
M^4$,
$\Str M^2$, and $\Str \bone$  conditions are not satisfied.
While all three of these supertrace
conditions are satisfied in SUSY theories,
spontaneous or soft SUSY-breaking preserves
only the $\Str \bone$ (and occasionally the $\Str M^2$)
condition at tree level \cite{SUSYreview}.

Let us now consider the corresponding situation in string theory.
A priori, there are three fundamental differences.  The first is that
in string theory, there are an {\it infinite}\/ number of states;  these
generally appear in towers whose levels are integer-spaced (in Planck-scale
units), and whose state degeneracies grow exponentially with mass.
This is why a regulator such as that in Eq.~(\ref{regulator}) must be chosen.
The second difference is that whereas field-theoretic states are characterized
by a single mass $M_i$, in string theory the energy of each state is described
through {\it two}\/ such quantities,
the separate left- and right-moving mass contributions
$M_i^{(L)}$ and $M_i^{(R)}$
whose squares always differ by integers.
A state is deemed ``physical'' if $M_i^{(L)}=M_i^{(R)}$, and ``unphysical''
otherwise;  note
that only the physical string states correspond to actual particles in
spacetime.  Nevertheless,
both types of states contribute to the string-theoretic one-loop
cosmological constant $\Lambda_\strin$.
Indeed, in string theory, $\Lambda_\strin$ is given by
\beq
     \Lambda_\strin ~\equiv~ \int_{\cal F} {d^2\tau\over ({\rm Im}\,\tau)^2}~
    Z(\tau)
\label{lambdadef}
\eeq
where the integration variable $\tau$ is the torus complex modular parameter,
and where the string partition function $Z(\tau)$ is a trace over the
Fock space of physical and unphysical string states,
\beq
    Z(\tau) ~=~ ({\rm Im}\,\tau)^{1-D/2}\, \sum_{\rm states} (-1)^F \,
         q^{[M_{i}^{(L)}]^2}
         \,\overline{q}^{ [M_{i}^{(R)}]^2}~
\label{Zstring}
\eeq
with $q\equiv e^{2\pi i\tau}$
and with all masses in units of the Planck mass.
In the usual string formulation, the modular invariance of $Z(\tau)$ allows
one to truncate the region of $\tau$-integration, as in Eq.~(\ref{lambdadef}),
to the fundamental domain of the modular group,
$      {\cal F}\equiv \left\lbrace
     \tau \,: \, |\tau|^2 \geq 1,
            {\rm Im}\,\tau >0,
              |{\rm Re}\,\tau| \leq 1/2 \right\rbrace $.

Since $\tau_2\equiv {\rm Im}\,\tau$ in string theory plays the role of the
Schwinger
proper time $t$ in field theory, we see that the region
$\tau_2\to \infty$ corresponds to the infrared, and $\tau_2\to 0$
to the ultraviolet.
Infrared divergences will thus be absent, as in field theory,
if there are no physical tachyonic states with $[M_i^{(L)}]^2=[M_i^{(R)}]^2<0$.
Indeed, this is part of what defines a physically consistent string theory.
Turning to the ultraviolet, however,
we see that the truncation of the region of $\tau$-integration to the
fundamental domain ${\cal F}$ has
already excluded the region near $\tau_2\to 0$.
This is the root of the well-known remarkable ultraviolet finiteness
properties of string theory, and the symmetry by which this occurs
(namely, modular invariance) is also part of what defines a
consistent string theory.

However, it is this feature which represents the third fundamental difference
between string theory and field theory, for we see that
the ultraviolet finiteness of $\Lambda_\strin$ has automatically arisen
through a truncation in the range of integration.
What we require, however, is an alternative
expression for $\Lambda_\strin$ whose finiteness explicitly
rests on the behavior of the string spectrum.  Indeed, it is only in this
way that we can exploit the finiteness of $\Lambda_\strin$ to derive a series
of supertrace mass formulas for string theory just as exist in field theory.
Fortunately, for a large class of tachyon-free string theories,
such an alternative expression exists \cite{kutsei}:
\beq
   \Lambda_\strin ~=~ {\pi \over 3} \, \lim_{\tau_2\to 0} \,
         \int_{-1/2}^{1/2} d\tau_1~ Z(\tau)~
\eeq
where $\tau_1\equiv {\rm Re}\,\tau$.
This class includes all unitary non-critical strings, critical Type-II
strings, as well as the phenomenologically
interesting case of $D>2$ critical heterotic strings.\footnote{
     Note that the critical heterotic case is special due
     to the appearance of unphysical tachyons with $[M^{(L)}]^2= -1$,
     $[M^{(R)}]^2=0$.
     However, for $D>2$, one can circumvent this difficulty by compactifying to
     a box
     of $(D-2)$-dimensional volume $V$, and taking $V\to\infty$ after
     the calculations are performed.  See Ref.~\cite{kutsei} for
     further details.  We thank D. Kutasov for discussions on this point.  }
Substituting the form of the string partition function $Z(\tau)$ in
Eq.~(\ref{Zstring}) and
explicitly performing the $\tau_1$ integral, we then obtain
\beq
   \Lambda_\strin ~=~ {\pi \over 3} \, \lim_{\tau_2\to 0} \,
     (\tau_2)^{1-D/2} ~ \sum_{\rm states} (-1)^F \,
         e^{-4\pi \tau_2 [M_i^{(L)}]^2 } ~ \delta_{M_i^{(L)},M_i^{(R)}}~.
\eeq
We thus see that in this formulation, only the masses of
the {\it physical}\/ string states are relevant.  Defining
$M_i\equiv M_i^{(L)}=M_i^{(R)}$, we therefore have
\beq
   \Lambda_\strin ~=~ {\pi \over 3} \, \lim_{\tau_2\to 0} \,
     (\tau_2)^{1-D/2}\, \sum_{\rm{phys.}\atop{states}} (-1)^F \,
        e^{-4\pi \tau_2 M_i^2 } ~,
\eeq
so that $\Lambda_\strin$ is indeed free of ultraviolet divergences if
and only if, as $\tau_2\to 0$,
\beq
   \sum_{\rm{phys.}\atop{states}} (-1)^F \,e^{-4\pi \tau_2 M_i^2 } ~\sim~
    (\tau_2)^\alpha ~~~{\rm with}~~ \alpha \geq  D/2-1~.
\label{conditiontwo}
\eeq

At this point we have shown that the masses of physical states throughout
the spectra of such consistent tachyon-free string theories must always
arrange themselves so as to satisfy Eq.~(\ref{conditiontwo}).
This result, however, is completely analogous to the corresponding
field-theoretic result
in Eq.~(\ref{conditionone}), and the different powers of $t$ or $\tau_2$ which
appear on the right sides of these equations reflect the extra finiteness
properties of string theory relative to field theory
(with the quartic and quadratic divergences in field theory
corresponding respectively to logarithmic divergences and
constant terms in string theory).
It is therefore tempting to proceed as for the field-theoretic
case, and expand the exponential to obtain the corresponding
supertraces.  However, in the string case
it is not technically proper to expand the exponential before taking the
limit, since our Fock space of string states is infinite-dimensional.
Rather, rigorously defining our string supertraces as in Eq.~(\ref{regulator})
and identifying $\gamma=4\pi \tau_2$, we should properly evaluate
these supertraces {\it without}\/ expanding the exponentials, but
rather by taking derivatives with respect to $\tau_2$ only {\it after}\/
the summation is performed:
\beqn
      \Str M^{2\beta}  &=&
    \lim_{\tau_2\to 0} \,\left\lbrace
    \left( {-1\over 4\pi}  {d\over d\tau_2}\right)^\beta
   \sum_i \,(-1)^F \, e^{-4\pi \tau_2 M_i^2} \right\rbrace \nonumber\\
     &=&
    \lim_{\tau_2\to 0} \,\left\lbrace
    \left( {-1\over 4\pi}  {d\over d\tau_2}\right)^\beta
    \left\lbrack {3\over \pi}\,\Lambda_\strin~ {\tau_2}^{D/2-1} \right\rbrack
             \right\rbrace.
\label{derivativemethod}
\eeqn
This yields, however,
the same results as we would have obtained
by expanding the exponentials.
In particular, we find from Eq.~(\ref{derivativemethod}) that
for general $D$,
\beq
      \Str\,M^{2\beta}~=~ 0~~~{\rm for}~~\beta< D/2-1, ~~\beta\in {\IZ}~,
\label{generalD}
\eeq
while for even $D$ we also have the result
\beq
      \Str M^{D-2}  ~=~ {3\over \pi}\, {(D/2-1)!\over (-4\pi)^{D/2-1}}\,
\Lambda_\strin ~.
\label{generalDlastone}
\eeq
Thus, for $D=4$, we find that the spectra
of all consistent unitary non-critical strings and critical
Type-II and heterotic strings
must satisfy the supertrace constraints in Eq.~(\ref{supertraces}).

We emphasize that our derivation has exploited only the fundamental
characteristics of string consistency --- namely,
the absence of physical tachyons and the existence of modular invariance.
We have {\it not}\/ imposed the finiteness of $\Lambda_\strin$
as is done in field theory for $\Lambda_\field$;  rather, the finiteness
of $\Lambda_\strin$ is a {\it consequence}\/ of these more fundamental
properties.  Hence the string case differs quite markedly
from the field-theoretic case.  For example, while
the vast majority of non-SUSY field theories do not obey any sort
of supertrace conditions, we see that it is generally {\it impossible}\/ to
avoid these constraints in string theory.  They are indeed generic
properties of the moduli space of such tree-level non-SUSY string vacua.

Given these
results, let us now discuss how string theory
manages to evade the phenomenologically undesirable consequences which
would arise in field theory.
It is here that the existence of an {\it infinite}\/ number
of string states proves crucial.
Indeed, it has recently been shown \cite{misaligned}
that the spectrum of any consistent string theory which is
modular invariant and free of physical tachyons will necessarily
exhibit a so-called ``misaligned SUSY''.
In the case of non-SUSY strings,
this hidden symmetry takes the form of a subtle
boson/fermion {\it oscillation}\/
in which, for example, any surplus of bosons at any
given string level necessarily implies a larger surplus
of fermions at a higher level, which in turn implies an even larger boson
surplus at an even higher level, and so forth throughout the
infinite tower of states.
Such behavior is sketched in Fig.~1 for
a simple string theory containing two sectors, a bosonic sector
with states at integer levels $M^2$ (in units of the Planck mass $M_0^2$),
and a
fermionic sector with states at levels $M^2\in {\IZ}+1/2$.
The numbers $g_M$ of bosonic minus fermionic states at each level
$M$ are indicated by the solid dots.
Although these two sectors are ``misaligned'' by a half-unit of energy, the
number
of such states at each level always grows exponentially according to
complicated
functions $\Phi(M)$ which are exactly equal and opposite for the two sectors.
Even for string theories containing {\it many}\/ sectors,
the sum of the corresponding functional forms $\sum_i\Phi_i(M)$
over all string sectors must always cancel,
and similar oscillations will appear.
Further details behind this ``misaligned SUSY''
can be found in Ref.~\cite{misaligned}.

\input epsf
\begin{figure}[t]
\centerline{\epsfxsize 4.0 truein \epsfbox {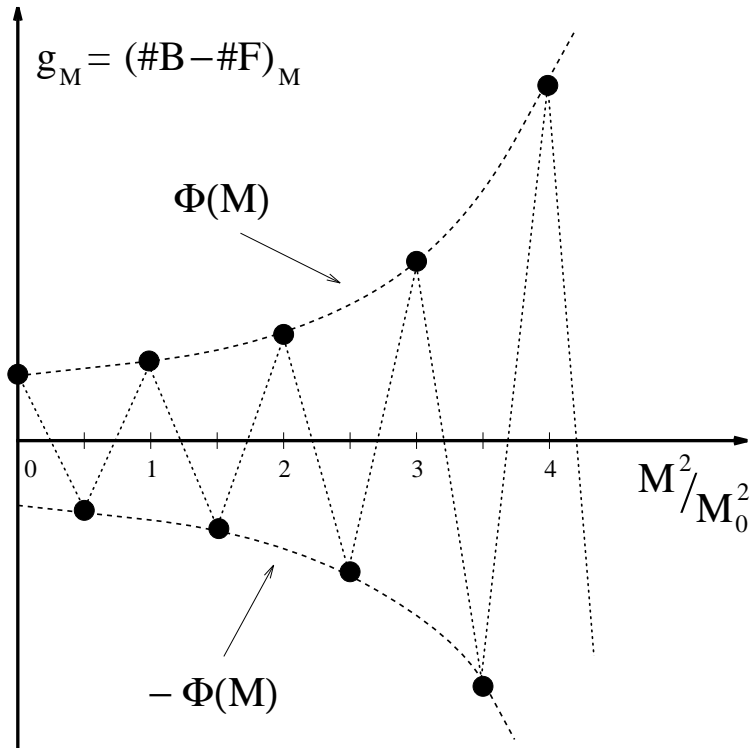}}
\nobreak
\caption{Misaligned SUSY and the resulting boson/fermion oscillations.}
\end{figure}

It is easy to check that degeneracies $g_M$ which behave in this
oscillatory fashion
will yield vanishing $\Str \bone$ when regulated as in Eq.~(\ref{regulator}).
This type of stringy boson/fermion oscillation is therefore precisely what
enables a string spectrum with exponentially growing numbers
of string states  to satisfy our supertrace constraints.
Moreover, since such oscillations achieve cancellations between
states at {\it different}\/ energy levels across the infinite string spectrum,
no strict multiplet-by-multiplet cancellations are necessary or even occur.
Indeed, it is possible to construct consistent non-SUSY string models
whose massless (observable) states are those of the Standard
Model, but whose (broken) superpartners are either absent
or at the Planck scale \cite{current}.  The spectra of such theories will
nevertheless satisfy our supertrace constraints.

We conclude with some final comments.  First, we observe that this
``misaligned SUSY'' mechanism can even be applied
directly in field theory, since this string-inspired oscillation
scenario does not depend on the particular scale $M_0^2$ of the level-spacing.
Indeed, all that is required for the absence of such divergences is
the cancellation of the degeneracy functions $\Phi(M)$.
Thus, for example, it may be possible to exploit
this mechanism to build an alternate {\it non-supersymmetric}\/ solution
to the gauge hierarchy problem.

Second, motivated by these supertrace results, we may ask whether there exist
non-SUSY string theories which nevertheless have vanishing $\Lambda_\strin$.
Indeed, such points in string moduli space would lead to string spectra
satisfying
both $\Str \bone = \Str M^2 =0$,
thereby ensuring at most logarithmic divergences
in any field theory containing the same numbers and energy distribution
of bosonic and fermionic states.  Furthermore,
such points would have vanishing dilaton one-point functions,
as required for vacuum stability at one loop and finite string amplitudes at
higher loops.
Unfortunately, despite various efforts \cite{gal}, no non-SUSY
models with vanishing $\Lambda_\strin$ have yet been constructed.
There do exist, however, string-like partition functions
$Z(\tau)$ which are non-vanishing ({\it i.e.}, non-supersymmetric),
but whose one-loop integrals $\Lambda$ vanish exactly \cite{newpf}.
Thus, there exist known non-supersymmetric distributions $\lbrace g_M\rbrace$
of bosonic and fermionic states which lead to vanishing $\Lambda_\strin$, and
for which both $\Str \bone$ and $\Str M^2$ cancel non-trivially.
Moreover, we see from Eq.~(\ref{generalD}) that
one can also obtain such $\lbrace g_M \rbrace$
with vanishing supertraces by considering non-SUSY strings
in higher dimensions.  For example, the degeneracies $\lbrace g_M\rbrace$
from the $D=10$ non-SUSY tachyon-free $SO(16)\otimes SO(16)$ string
have ${\rm Str}\,{\bf 1}$, ${\rm Str}\,{M^2}$,
${\rm Str}\,{M^4}$,  and ${\rm Str}\,{M^6}$
all vanishing.

Third, we emphasize that our definition of
the string-theoretic supertraces in Eq.~(\ref{regulator})
is rooted in the actual string spectrum, and
realizes the supertrace as an explicit sum over string states.
As such it is completely general, and applies to large classes of tachyon-free
four-dimensional string theories.
By contrast, alternate supertrace calculations
\cite{antoniadis} consider only a particular family of non-SUSY
string vacua which are continuously connected to a supersymmetric point,
and define the supertraces through an expansion of $\Lambda_\strin$
with respect to the relevant SUSY-breaking parameter.
Understanding the relation between these two approaches is an important issue.

Finally, we point out that an outstanding problem
in string theory has been to understand the origins
of misaligned SUSY as a symmetry, and to determine the kinds of dynamical
SUSY-breaking scenarios which lead to such boson/fermion oscillations.
Our results concerning the supertrace implications of misaligned SUSY
will therefore be a useful tool in this quest.

  \bigskip
  \medskip
\leftline{\large \bf Acknowledgments}
  \medskip
  \smallskip
One of us (KRD) thanks I. Antoniadis, C. Bachas,
S. Chaudhuri, E. Kiritsis, C. Kounnas, D. Kutasov, C.S. Lam, J. Louis,
and F. Zwirner for discussions, and the Technion ITP for hospitality during
a visit when portions of this paper were written.
RCM also thanks C. Burgess for discussions.
This work was supported in part by DOE Grant No.\ DE-FG-0290ER40542,
the US/Israel Bi-National Science Foundation, the Technion VPR Fund,
NSERC (Canada), and FCAR (Qu\'ebec).
  \bigskip
  \bigskip
  \medskip


\bibliographystyle{unsrt}

\vfill\eject
\end{document}